\documentclass[journal]{IEEEtran}
\usepackage[T1]{fontenc}
\usepackage[latin9]{inputenc}
\usepackage{amsmath}
\usepackage{amssymb}
\usepackage{amsthm,bm}
\usepackage{url}

\theoremstyle{plain}

\theoremstyle{remark}

\makeatother

\usepackage{babel,verbatim,balance}
\usepackage[rgb]{xcolor}
\usepackage{graphics, graphicx}
\usepackage{acronym}
\usepackage{upgreek}
\usepackage{setspace,gensymb}

\newcommand{\change}[1]{{\color{black}{#1}}}
 
\acrodef{bs}[BS]{base station}
\acrodef{csi}[CSI]{channel state information}
\acrodef{tdd}[TDD]{time division duplexing}
\acrodef{ris}[RIS]{reconfigurable intelligent surface}
\acrodef{rss}[RSS]{received signal strength}
\acrodef{toa}[ToA]{time of arrival}
\acrodef{poa}[PoA]{phase of arrival}
\acrodef{aoa}[AoA]{angle of arrival}
\acrodef{aod}[AoD]{angle of departure}
\acrodef{ula}[ULA]{uniform linear array}
\acrodef{snr}[SNR]{signal-to-noise ratio}
\acrodef{gdop}[GDOP]{geometric dilution of precision}
\acrodef{em}[EM]{electro-magnetic}
\acrodef{fim}[FIM]{Fisher information matrix}
\acrodef{los}[LoS]{line-of-sight}
\acrodef{nlos}[NLoS]{non-line-of-sight}
\acrodef{bs}[BS]{base station}
\acrodef{peb}[PEB]{position error bound}
\acrodef{uwb}[UWB]{ultra wideband}
\acrodef{slam}[SLAM]{simultaneous localization and mapping}
\acrodef{psd}[PSD]{power spectral density}
\acrodef{ofdm}[OFDM]{orthogonal frequency division multiplexing}
\acrodef{rt-tof}[RT-TOF]{round-trip-time-of-flight}
\acrodef{tdoa}[TDoA]{time-difference-of-arrival}
\acrodef{coa}[PoA]{phase of arrival}
\acrodef{mse}[MSE]{mean squared error}
\acrodef{slam}[SLAM]{simultaneous localization and mapping}
\acrodef{ue}[UE]{user equipment}

\begin{document}
 
\title{Reconfigurable, Intelligent, and Sustainable Wireless Environments for 6G Smart Connectivity}
\author{Emilio Calvanese Strinati, George C. Alexandropoulos, Henk Wymeersch, Beno\^{i}t Denis, Vincenzo Sciancalepore, Raffaele D'Errico, Antonio Clemente, Dinh-Thuy Phan-Huy, Elisabeth De Carvalho, and Petar Popovski
\thanks{Emilio Calvanese Strinati, Beno\^{i}t Denis, Raffaele D'Errico, and Antonio Clemente are with CEA-Leti; 

George C. Alexandropoulos is with the National and Kapodistrian University of Athens; 

Henk Wymeersch is with the Chalmers University of Technology; 

Vincenzo Sciancalepore is with NEC Laboratories Europe; 

Dinh-Thuy Phan-Huy is with Orange Labd Network; 

Elisabeth De Carvalho and Petar Popovski are with the Aalborg University.}
}

\maketitle
\begin{abstract}
Various visions on the forthcoming sixth Generation (6G) networks point towards flexible connect-and-compute technologies to support future innovative services and the corresponding use cases. 6G should be capable to accommodate ever-evolving and heterogeneous applications, future regulations, and diverse user-, service-, and location-based requirements. A key element towards building smart and energy sustainable wireless systems beyond 5G is the \emph{Reconfigurable Intelligent Surface (RIS)}, which offers programmable control and shaping of the wireless propagation environment. 

Capitalizing on this technology potential, in this article we introduce two new concepts: $i$) \emph{wireless environment as a service}, which leverages a novel RIS-empowered networking paradigm to trade off diverse, and usually conflicting, connectivity objectives; and $ii$) \emph{performance-boosted areas} enabled by RIS-based connectivity, representing competing service provisioning areas that are highly spatially and temporally focused. 
We discuss the key technological enablers and research challenges with the proposed networking paradigm, and highlight the potential profound role of RISs in the recent Open Radio Access Network (O-RAN) architecture.  
\end{abstract}

\section{Introduction}
\IEEEPARstart{T}{he} fifth Generation (5G) of communication networks, which is still at an early deployment stage, provides a single platform for a variety of services and vertical applications.
However, the design of novel enablers for beyond-5G connect-and-compute networks is already envisaged for better satisfaction of future needs, both individual and societal~\cite{CBG19}. 
While some of the solutions will be included in the long-term evolution of 5G, others will require more radical changes, leading to the standardization of the future 6G wireless networks.

The motivation for research on wireless connectivity beyond 5G emerges, similarly to what happened in the past, from  the combination of late technological readiness and the continuously evolving economic and societal challenges stemming from respective trends. A number of Key Performance Indicators (KPIs) has been recently discussed~\cite{CBG19}, including up to $10~ \text{Gb}/\text{s}/\text{m}^3$ capacity, $100$~$\mu s$ latency, $1$~Tb/J energy efficiency, and $1$~cm localization accuracy in $3$ Dimensions (3D). Depending on future application and service needs, these  KPIs will not all have to be achieved simultaneously and in all occasions, but rather locally in space and time, with a high degree of flexibility and adaptivity. This type of agility can be only achieved under the condition that the wireless propagation environment is not completely uncontrollable and hostile. 

In relation to this, recently there has been a surge of interest in  Reconfigurable Intelligent Surfaces (RISs)~\cite{Kaina_metasurfaces_2014} as hardware-efficient and highly scalable means to realize desired dynamic transformations on the propagation environment in wireless communications~\cite{huang2019reconfigurable}. RISs are human-made surfaces with hundreds or even thousands of nearly passive (reconfigurable) elements, which can support various functionalities, such as signal relaying/nulling, range/position estimation, obstacles/activity detection, pencil-like beamforming, and multipath shaping. They are suitable for controlling radio wave propagation~\cite{huang2019reconfigurable} and the geometry in multipath-aided localization~\cite{WD20}, limiting ElectroMagnetic Field (EMF) exposure, mitigating obstructions, and extending radio coverage in dead zones. The RIS technology can be used for coating objects in the environment~\cite{huang2019holographic} (e.g., walls, ceilings, mirrors, or appliances), and can operate as a reconfigurable reflector, or as a transceiver for massive access~\cite{JSAC2020_risma} when equipped with active Radio-Frequency (RF) elements.

In this article, we introduce a novel wireless connectivity paradigm comprising negligible-power RISs and conventional network nodes. This paradigm aims at jointly optimizing the radio wave propagation environment with the existing network infrastructure to realize highly concentrated (i.e., selective in time and space) service provisioning to intended end-users, while removing energy from regions where accidental or non-intended users are present. We coin the concept of the \emph{Wireless Environment as a Service (WEaaS)}, which offers dynamic radio wave control for trading-off high capacity communications, energy efficiency, localization accuracy, and secrecy guarantees over eavesdroppers, while accommodating specific regulations on spectrum usage and restrained EMF emissions. 
\change{We discuss the core architectural components of the proposed Reconfigurable, Intelligent, and Sustainable wireless Environments (RISE) for 6G connectivity, while validating their openness features within the already defined O-RAN-compliant landscape and, in turn, enabling the concept of the spatially-focused} \textit{Performance-Boosted Areas (PBAs)}. The main challenges and open research directions for the envisioned smart wireless connectivity paradigm are discussed.

\section{The RISE Network Paradigm}
The RISE network paradigm integrates multiple spatially distributed RISs within the 5G-and-beyond network components. The RISs can be heterogeneous in terms of size and low-power hardware technologies. This approach fosters an unprecedented reuse of the mobile network infrastructure, leading to reduced operational/capital expenditure, but also lower energy consumption through highly spatially focused service provisioning. 

\subsection{Core Network Components}
\begin{figure}
     \centering
     \includegraphics[width=\columnwidth]{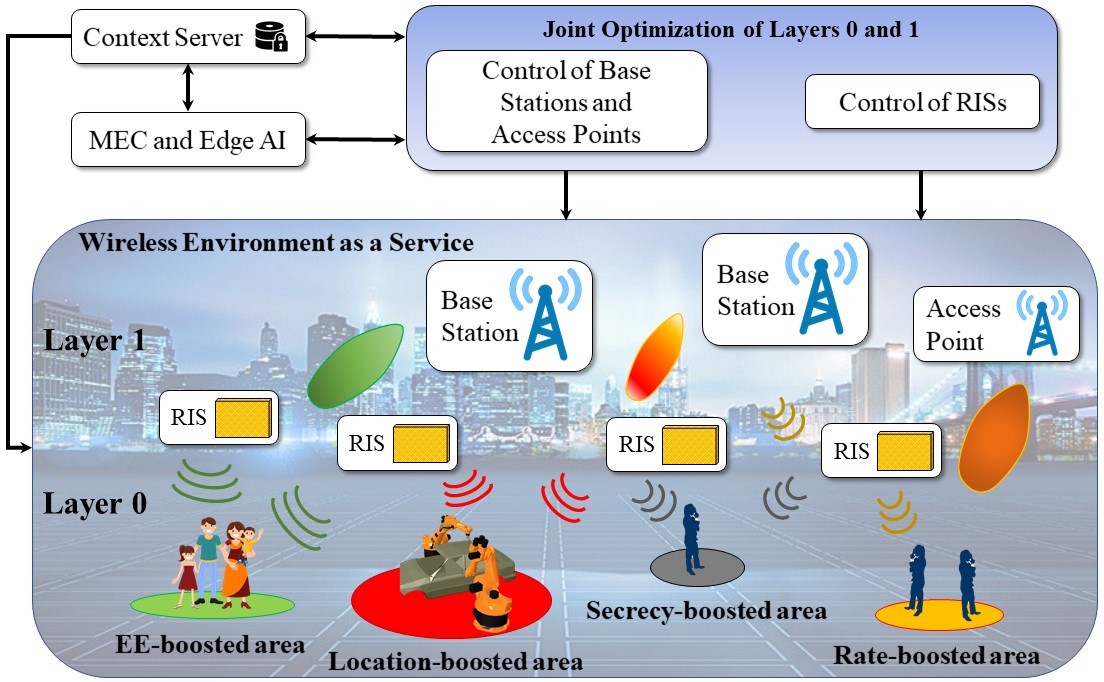}
     \caption{Core components of the proposed RISE network paradigm.}
    \vspace{-0.1cm}		
     \label{fig:conceptRISE6G}
 \end{figure}

The core components of the proposed networking paradigm are illustrated in Fig$.$~\ref{fig:conceptRISE6G}. The novel Layer-$0$ of the Smart Radio Environment (SRE) including multiple and possibly densely deployed low-cost and nearly passive RISs is introduced, which will serve as the programmable wireless medium where the physical Layer-$1$ of the current and future wireless standards will operate. The RISs can be coated in objects and appliances in the environment in a highly scalable and potentially cell-free and/or multi-tenant manner. As depicted in Fig$.$~\ref{fig:conceptRISE6G}, the proposed paradigm deploys an orchestration module responsible for the joint optimization of the available hardware resources in the bottom two layers (i.e., all RISs and conventional network transceivers) in conjunction with the context and localization server, and the network entities that host the Multi-Access Edge Computing (MEC) and edge Artificial Intelligence (AI) functionalities. 

The low-latency as well as the reliable AI and machine learning services at the edge of the RISE network will enable optimal resource allocation schemes to be learnt over time, with the aim of ensuring energy efficiency, low EMF, increased secrecy spectral efficiency at Layer-$1$, and communicate-and-compute services under strict time constraints. The AI algorithms will be also intended for providing automation and self-healing orchestration capabilities for monitoring and optimal operational decision, in cases of dynamic network changes or failures. In this way, a dynamic scheduler will learn how to orchestrate the system resources, while deciding on-the-fly which users should be served through which of the available RISs. User mobility patterns can be also incorporated and/or learned when necessary, in order to enhance the cognition and adaptation capabilities of the proposed RISE network paradigm. 

\subsection{RISs Control}
The RISs can be seen as the key Layer-$0$ technology, affecting the radio wave propagation environment on top of which Layer-$1$ data and control signals are transmitted. 
There are two principal ways to control an RIS:
\begin{itemize}
    \item \emph{In-band control}, where the properties of the RIS are dynamically configured by using the wireless signals whose propagation is supposed to be affected by the RIS. The key feature of in-band control is that the RIS should be reconfigured under the constraint that it keeps the control channel operational.
    \item \emph{Out-of-band control}, where the properties of the RIS are controlled through a communication interface that is slightly or not at all affected by the reconfiguration of the RIS. In this case, the control channel is decoupled from the RIS reconfiguration. 
\end{itemize}
The in-band control can be explicit or implicit. In \emph{explicit in-band control}, the RIS contains a full radio interface, such as 5G New Radio (NR) or a potential 6G interface, that is capable to exchange RIS-related control signals. Explicit in-band control is necessary when the RIS needs to be shared among competing stakeholders, e.g., several service providers in the same area, that need to be able to make contractual requests to the RIS. In \emph{implicit in-band control}, the RIS reconfiguration relies on the fact that the RIS can sense and interpret the received wireless signals. This means that the RIS implementation exhibits sensing capabilities \cite{huang2019holographic}. Implicit in-band control can be used when the RIS supports PBAs and changes the environment at the benefit of multiple stakeholders that deploy or purchase wireless service in that area. Based on these two main ways to control an RIS, various hybrid and composite control schemes can be designed. 

Another important dimension for making a taxonomy of the Layer-$0$ reconfiguration describes the elementary resources that can be controlled and allocated to a stakeholder. In general, these resource can be space (geometrical parts of the RIS), time (time intervals in which the RIS is allocated to a given stakeholder), frequency (for which band is the RIS configuration valid), or a combination thereof. 

\subsection{Performance Benefits}
\begin{figure}
     \centering
     \includegraphics[width=\columnwidth]{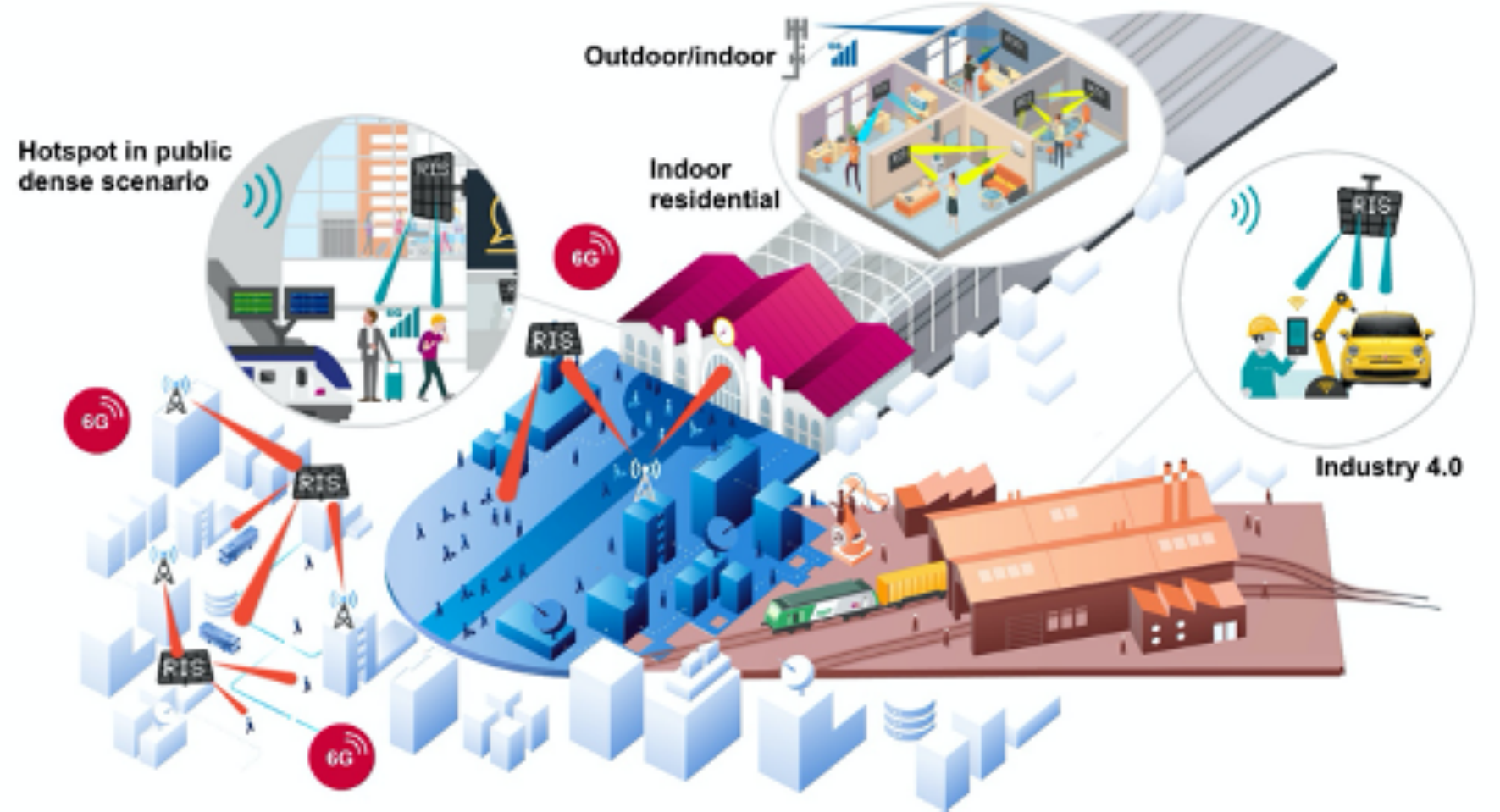}
     \caption{
     Applications for the proposed RISE wireless connectivity paradigm, enabled by a plurality of power-efficient RISs integrated with the conventional network infrastructure.}\vspace{-0.1cm}
     \label{fig:RISE6GEnvironment}
 \end{figure}

The envisioned RISE network paradigm enables performance-boosted wireless connectivity, as shown in the example depicted in Fig$.$~\ref{fig:RISE6GEnvironment}. As demonstrated, multiple RISs can be deployed outdoors, in indoor hot spots, and in public user-dense scenarios (e.g., metro/train stations, airports, and shopping malls), in indoor residential scenarios to boost both indoor and outdoor-to-indoor wireless communications, as well as in vertical scenarios (e.g., the Industry 4.0). In such operational contexts, making use of nearly passive and low-cost network infrastructure instead of deploying additional active hot spots, will result in avoidance and alleviation of several network installation and maintenance issues. Hence, substantial gains are expected in terms of achieving: \textit{i}) delay minimization due to the avoidance of long site negotiation; \textit{ii}) reduced energy consumption for operating the dense network of cell-free and/or multi-tenant RISs, as compared with conventional transceivers and relays; \textit{iii}) avoidance of high data rate wired links for controlling the RISs via efficient control schemes realized wirelessly; and \textit{iv}) minimization of the cost and effort for the installation and de-installation of radio access infrastructure.
State-of-the-art and future visions in RIS fabrications~\cite{Kaina_metasurfaces_2014,HFL19} witness that they will be lightweight and aesthetically transparent devices, which can be deployed near the customers. This unique feature will allow consideration of a plethora of existing power-plugs in urban areas as RIS-coated network nodes (e.g., glass-made building facades, billboards, publicity displays, walls with plugs near seated customers in coffee shops, offices, or waiting rooms).

\section{The RISE-Empowered Concepts}
Compared to business-as-usual, where communications must be optimized given an imposed and uncontrolled wireless medium, the envisioned SREs will enable the granting of highly localized quality of experience and specific service types to the end users. This leads to the two novel concepts brought by this article: \textit{WEaaS} and \textit{PBAs}. The proposed RISE network paradigm aims at going beyond the classical 5G use cases, which require that the network is tuned to one of the available service modes 
in an orthogonally isolated manner, offering non-focalized areas of harmonized and balanced performance. The PBAs, however, are defined as dynamically designed regions that can be highly localized, {offering customized high  resolution manipulation of radio wave propagation }to meet selected KPIs.

\subsection{Performance-Boosted Areas}
Figure~\ref{fig:RISE6GEnvironment} illustrates extreme data rates in a rate-boosted area (e.g., the highly localized hotspot for infotainment in the train station), as well as accurate localization and radio mapping in \textit{localization-boosted area}
(e.g., to precisely position User Equipment (UE), assets, or goods in the Industry 4.0 factory/warehouse). 
Similarly, a \textit{low-EMF area} will enable dynamic adaptation of EMF exposure to end users according to local regulations, while an \textit{energy-efficient-boosted area} will ensure targeted performance of wireless connectivity with low energy consumption. {A secrecy-boosted area will operate with improved secrecy and security guarantees (e.g., professional office environments) by at least boosting the level of signal over intended users and at best reducing the level of signal over eavesdroppers. The latter strategy for the eavesdropper faces a research challenge that is well recognized in the cognitive radio domain: the lack of knowledge regarding this user. Instead, we believe that RIS-aided localization and positioning of non-intended users will be potentially helpful in this context.} In the proposed RISE network paradigm and envisioned applications, the RIS technology enables a cost-effective adaptation of the wireless environment, allowing the aforementioned KPIs to be fulfilled simultaneously and concurrently within a spatially small environment.  

\begin{figure}
     \centering
     \includegraphics[width=0.98\columnwidth]{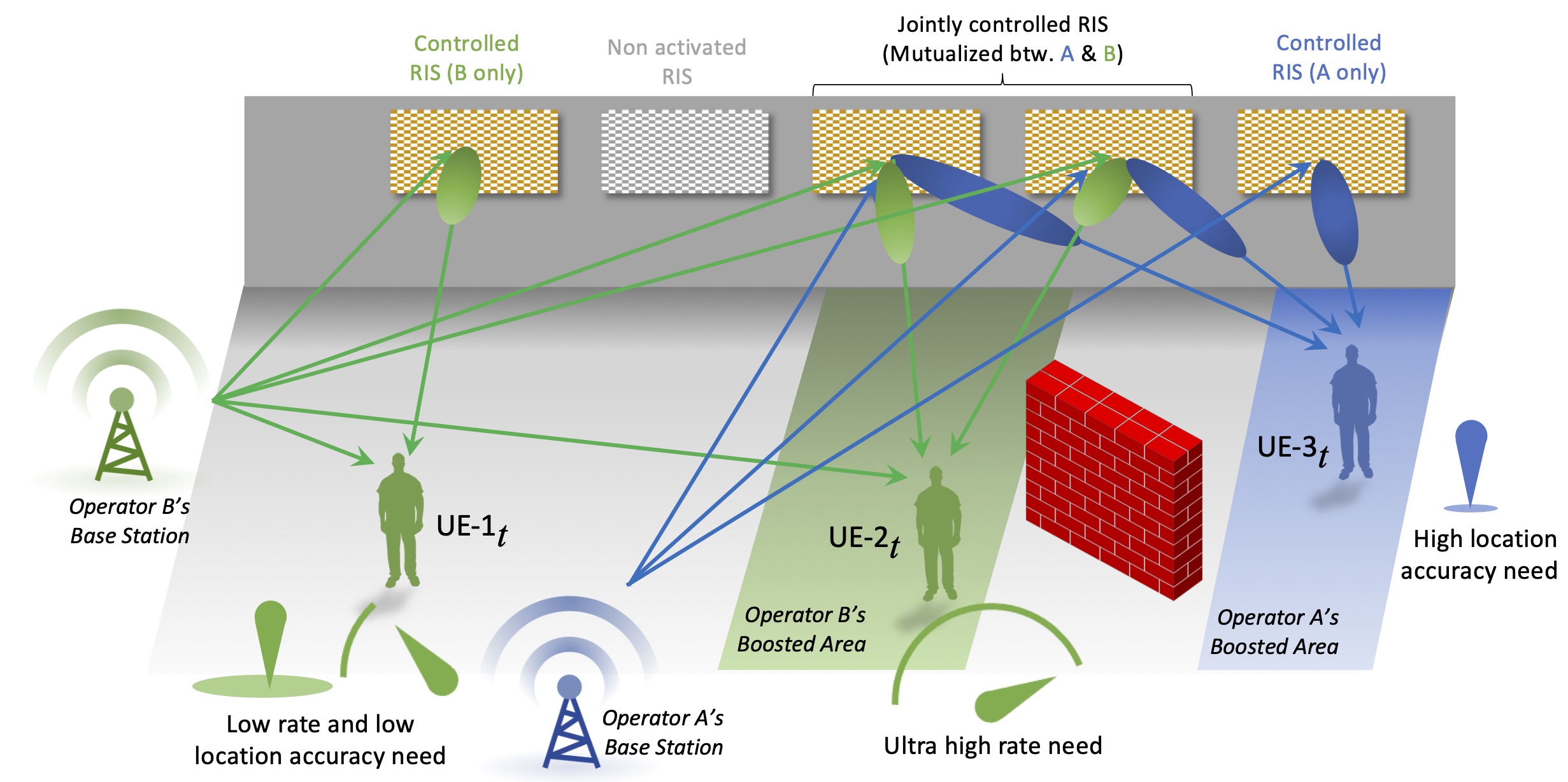}
     \caption{
     Example of an RIS-empowered multi-user $2$-operator WEaaS case, where distinct service levels with service continuity are delivered by two base stations (blue and green), based on the respective positions occupied by their served UEs (e.g., inside or outside operator-wise boosted areas), as close as possible to their actual needs. 
     }\vspace{-0.1cm}
     \label{fig:Boosted_Perf_&_Service_Continuity_Example}
 \end{figure}   


\subsection{Multi-Operator RIS WEaaS}
\change{
Figure~\ref{fig:Boosted_Perf_&_Service_Continuity_Example} details an example of the RIS-empowered WEaaS concept in a use case where two distinct operators cooperate to jointly control the same RISs in order to satisfy the communication needs (e.g., data rates and localization accuracy) of their UEs inside and outside their respective boosted areas. For this purpose, the operators may rent a subset of the pre-deployed RISs in two principal ways: $i$) orthogonal, via reserving a certain sub-area of each RIS; or $ii$) non orthogonal, where joint use of the whole RIS area is allowed and the RIS resources are allocated through spatial tuning according to the policies set by the RIS owner. It is, however, rather difficult to address different frequencies based on interleaved RIS elements and/or optimize the phase distributions over multiple frequencies with no perturbation. Hence, this multi-user multi-operator communication scenario can be naturally formulated as a multi-objective multi-constraint optimization problem where the constraints can be based on, e.g., decoding objectives and frequency dependency. This would end up into novel and non-trivial forms of broadcast communications with shared RISs. 

In Fig$.$~\ref{fig:SR_Results}, considering Rayleigh fading channels for all links, we demonstrate the ergodic sum-rate performance of a $2$-operator WEaaS case with $2$ single-antenna UEs and a shared RIS, and compare it with that of two individually controlled RISs, and with the case where no RIS is deployed. For reduced RIS control overhead and respective energy consumption, we have assumed that the RISs use random phase configuration per channel usage. In the case of the individual RISs, each $8$-antenna Base Station (BS) selfishly adopts Maximal Ratio Transmission (MRT). In the shared RIS case, we have considered both individual MRTs on the distinct compound direct and BS-RIS-UE channels, as well as coordinated beamfoming at the BSs via the Regularized Zero Forcing (RZF) precoding scheme. The latter schemes were analogously considered when RISs were absent. The performance curves in the figure indicate that the WEaaS case can provide substantial link budget gains compared to the case where two RISs are deployed individually, and this happens with half of the hardware cost and power consumption.

Multipath localization capabilities can be harnessed using a single BS in the presence of obstacles. As shown in Fig.~\ref{fig:Boosted_Perf_&_Service_Continuity_Example}, the uncertainty region associated with the UE's estimated location can be dynamically fine-tuned in terms of both size and orientation, according to \textit{a priori} location-dependent application requirements~\cite{WD20}. 
This RIS setup can fulfill the needs of a more demanding application (i.e., boosted regime or area), while possibly favoring accuracy in one particular dimension. 
This will be particularly relevant in challenging setups such as smart factories, where various levels of location-based authorization and/or safety must be guaranteed in specific zones occupied by machine operators, robots, or mobile assets, and under fast changing and heavily obstructed  radio propagation conditions. 
  \begin{figure}
     \centering
     \includegraphics[width=\columnwidth]{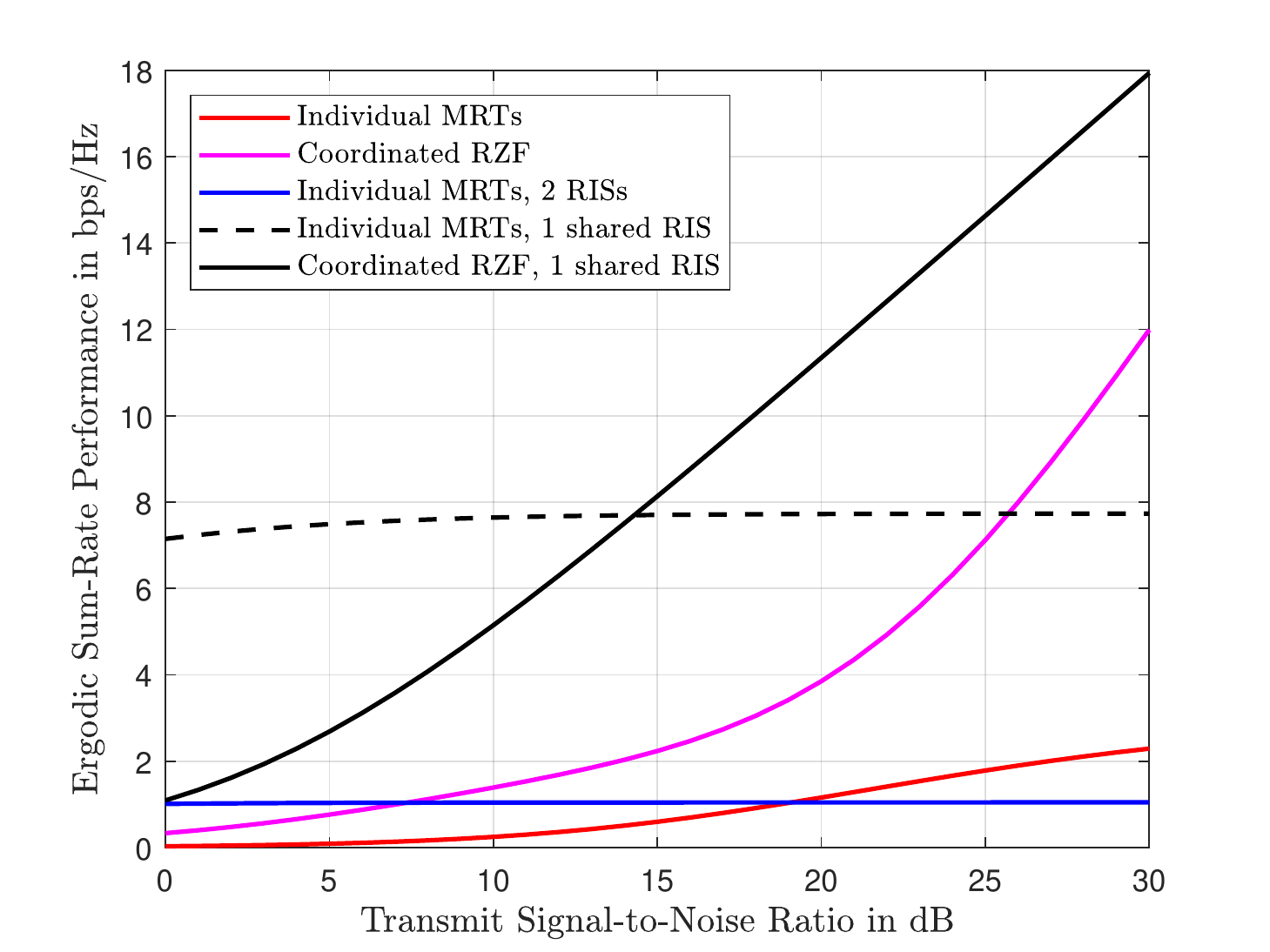}
     \caption{Ergodic sum rate of RIS-empowered $2$-user $2$-operator WEaaS with an $100$-element shared RIS. The performance with two individually controlled $100$-element RISs and without any RIS is also illustrated. The strengths of the BS-RIS and BS-UE channels are $30\%$ and $1\%$ percent, respectively, of those of the RIS-UE channels. 
     }
     \label{fig:SR_Results}
 \end{figure}}




\section{6G Technology Enablers and Challenges}
While building the RISE networking paradigm, a number of technological challenges will be encountered within the 6G journey. Hereafter, we detail the main ingredients to create a seamless and practical integration of the proposed WEaaS and PBA concepts into 6G wireless networks.

\subsection{RIS-Empowered Communication Models}
The design of practical communication systems has always been restricted to the optimization of transmission methods (e.g, coding, modulation, and beamforming), detection mechanisms (e.g., synchronization and channel estimation), and other operation protocols (e.g., retransmissions and resource allocation), conditioned on a certain wireless channel. The traditional goal of the communication system design is to optimize spectral efficiency and approach Shannon capacity, assuming that the environment is not controllable. Conversely, the proposed RISE network paradigm brings the communication technology into a new era, where the propagation channel itself can be controlled. This perspective will lead to novel mathematical models and derivations combining information theory for communications, ElectroMagnetic (EM) propagation, and reconfigurable metamaterials. 

To study the dynamic wave propagation control offered by RISs, realistic models that capture the impact of their changing states on the signal propagation are required. The current accepted assumption is conceiving RIS as a perfect mirror, while in reality different grating lobes can appear. Only recently, antenna theory was used to compute the electric field in the near- and far-field regions (distinguished by the Fraunhofer distance) of a finite-size RIS, and have proved that an RIS is capable of acting as an anomalous mirror in the near field of the array~\cite{garcia2019reconfigurable}. Considering plane wave propagation (i.e., far-field communications) at the millimeter wave band in~\cite{khawaja2020coverage}, the measured received power from passive reflectors was compared against ray tracing simulations. However, the reflections from the surrounding environment were not considered, which can change the current density within the RIS, thus modifying its local EM response. Furthermore, the interactions among reflecting RISs need to be analytically modeled in order to accurately characterize controllable radio wave propagation both in the near and far fields. 

\subsection{\change{Open and Flexible Network Architecture Design}}
\begin{figure*}
     \centering
     \includegraphics[width=\linewidth, clip]{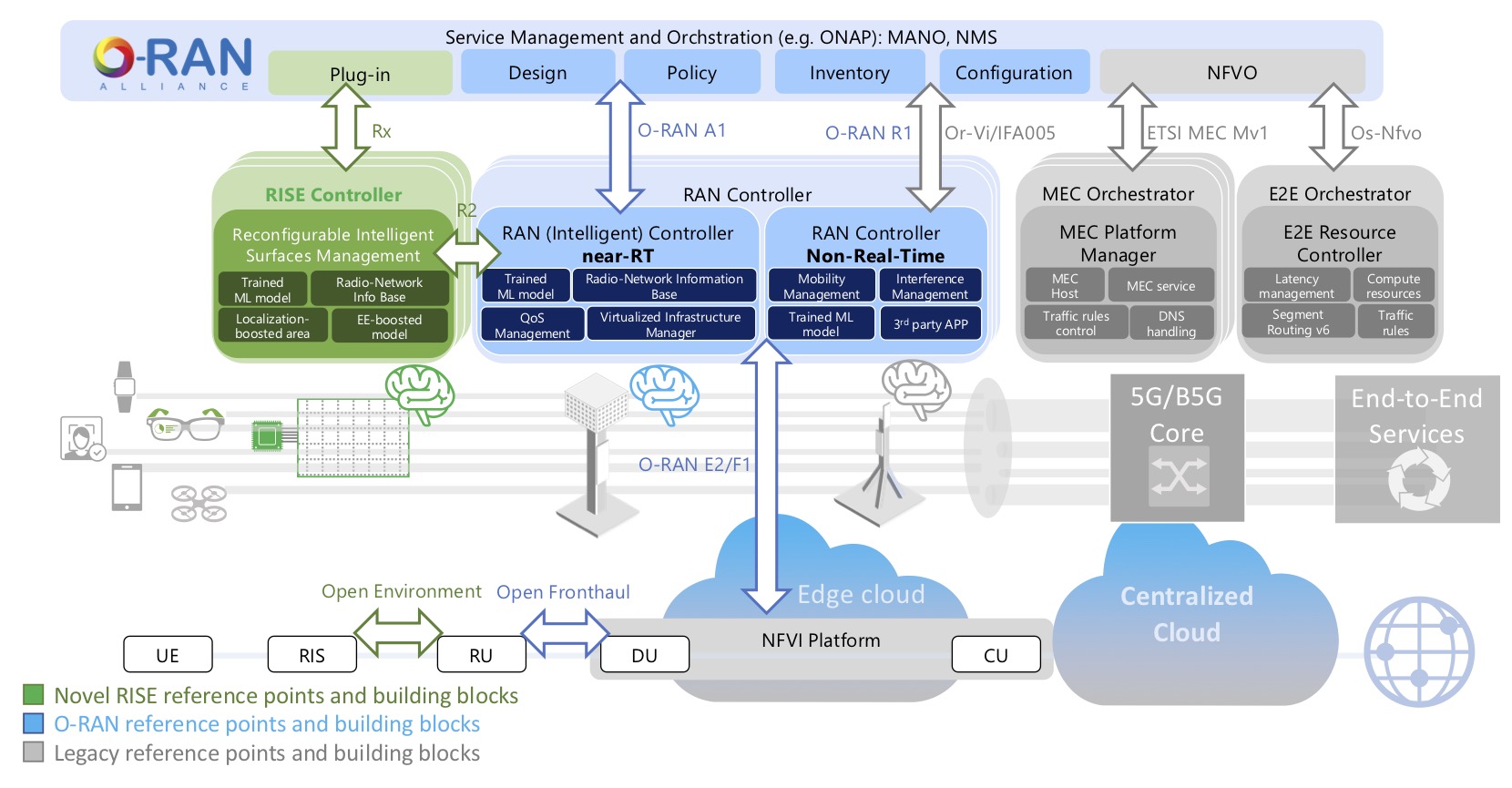}
     \caption{Proposed integration of the RISE-based network with O-RAN-compliant interfaces and legacy building blocks. \change{The new reference points Rx and R2 are included to interface the RISE controller,} while leaving open the interface between Radio Units (RUs) and RISs. DU, CU, and ML stand for Distributed Unit, Centralized Unit, and Machine Learning, respectively.}\vspace{-0.2cm}
     \label{fig:RISE6GArch}
 \end{figure*}
 
The huge efforts of the main network business players has recently led to a common concept for the 6G era: \emph{openness}. In this regard, the shared vision for the novel network architecture requires new interfaces, thereby opening the telecom industry's door to hardware and software providers. In particular, the network operators may freely decide to acquire general purpose hardware to provision their own Radio Access Networks (RAN) with customized virtual network functions following the Open-RAN paradigm, as per the official O-RAN Architecture Description v2.0. 
In this context, the deployment of RISs may increase the opportunity to create fully interoperable SREs that can further push this drastic openness change into the networking industry.

\change{The proposed RISE network components can be plugged into a shared O-RAN view via the NFV/MANO integration, as depicted in Fig.~\ref{fig:RISE6GArch}}. We envision a new network segment between the RAN elements and the final UEs, \change{wherein the elements of the RISs can be automatically commanded by the RISE controller within short timescales (e.g., milliseconds) to drive the overall network efficiency at the (extreme) edge. In particular, such a controller will be interfaced to the orchestration layer via a new reference point (Rx) that instructs transmitter configurations on long timescales (e.g., minutes) for specific environments or given use cases}. In parallel, the RISE controller can adjust the RIS parameters based on the received feedback as well as the desired KPIs. \change{Additionally, a new interface (R2) can be designed to create a direct link between the near-Real-Time (near-RT) RAN Intelligent Controller (RIC) and the RISE Controller. This will allow to simultaneously control the configurations of RIS and the beamforming of BSs in very short timescales.} 
Finally, an ``Open Environment'' interface can be developed between the RIS elements and the radio units handling the digital front, to bring flexibility and programmability into the WEaaS concept.

\subsection{Multi-Functional RIS Hardware}
Several technologies are being considered for RIS hardware fabrication, including reflectarrays, transmitarrays, and software-defined metasurfaces. If compared to classical phased arrays which require phase shifters and power amplifiers, RISs are nearly passive structures. They integrate switches and phase manipulating units, like PIN diodes, varactors, and liquid crystals, which are used for electronically controlling their phase shifts and impedance characteristics.

A reflective RIS operates as an EM mirror, where an incident wave is reflected in a desired direction with specific radiation and polarization characteristics. On the other hand, a transmissive RIS operates as a lens or a frequency selective surface, where the incident field is manipulated (e.g., filtering, polarization, and beam splitting) and re-transmitted. Although RISs have great potential to implement advanced EM wave manipulations, only simple functionalities, such as electronic beam steering and multi-beam scattering, have so far been demonstrated in the literature. Binary tunable reflective RISs were used in~\cite{Kaina_metasurfaces_2014} for indoor passive EMF control targeting improved coverage, while~\cite{HFL19} demonstrated that an RIS can be deployed for increasing the effective rank of an indoor channel matrix. Beamforming functionalities with a fabricated transmissive RISs have been recently demonstrated up to Ka-band~\cite{diaby20192} with a phase resolution up to $2$ bits. 
 
The cabling interconnection of the RIS unit elements for their dynamic control is still very complicated, even for sub-6 GHz wireless communication. This becomes much more challenging at higher frequencies. A possible solution would be to consider multi-core fiber optics for providing dense interconnections. While a variety of functionalities have been successfully presented with different RIS hardware technologies, the simultaneous demonstration, implementation, and control of multiple functionalities has not been yet reported. Furthermore, polarization control has only been partially studied. Finally, it is noted that current fabrication approaches are similar to those for conventional antennas. We believe that novel methodologies tailored to RISs need to be invented, which span operating frequencies from sub-6 GHz to mmWave, and the more exploratory sub-THz bands. Those methodologies should take into account the multi-function, energy efficiency, scalability, and flexibility requirements envisioned for future 6G networks. 
     
\subsection{RIS Integration Protocols}
To orchestrate the control of RISs, allocate in an optimized way the augmented radio resources that they offer, and establish interactions among RISs and the legacy network infrastructure, novel RIS integration protocols are needed. Of paramount importance to the WEaaS concept is the inference of the various wireless channel states that the different RIS configurations result in, as well as their dynamic management for realizing PBA-based smart connectivity.

Channel Acquisition and Tracking (CAT) in wireless networks incorporating RISs is a recent active area of research including techniques mainly falling into two categories~\cite{risTUTORIAL2020}. The first category assumes that the RISs are passive, thus realized with minimal hardware complexity as well as signal processing and power requirements, while the second category considers RISs that possess estimation and sensing capabilities. When passive RISs are deployed, CAT can be only handled by conventional receivers via dedicated channel sensing/estimation protocols. Those protocols involve sequences of activation patterns for the RIS unit elements or leverage the channel sparsity assumption in the angular domain to treat CAT as a compressed sensing problem. However, further research is required to reduce the CAT overhead (i.e., training period and amount of control signaling), especially in scenarios with mobile UEs. When an RIS is equipped with sensors or a single receive RF chain~\cite{hardware2020icassp}, it can perform CAT via AI techniques. Research on this category is still at an exploratory stage and requires significant progress to reach a mature stage for implementation. It is critical to account for the need for extra hardware depending on the considered RIS deployment application, and design methodologies for choosing which RIS should be capable of sensing and which not. For both CAT categories, the cell-free nature of in-band and out-band RISs should be taken under consideration to devise dynamic clustering algorithms for their efficient orchestration, as well as distributed optimization approaches with feature learning aspects for the multiple RISs and BSs. 

\subsection{Multi-Access Edge Computing and Learning}
The goal of MEC is to bring cloud functionalities at the edge of wireless networks. One of the most impactful MEC services is the offloading of computations from mobile UEs and other end terminals to nearby MEC servers, with the aim of saving device energy or running sophisticated applications with low energy and latency constraints. Computation-offloading problems can be classified as either static or dynamic strategies. The static formulation deals with short-time applications, in which mobile users send a single computation request, specifying also a service time. Conversely, in a dynamic scenario, the user application continuously generates data to be processed, often with an unknown rate. 

Some preliminary schemes for computation offloading leveraging the dynamic wave propagation control offered by RISs have appeared lately (e.g.,~\cite{liu2020intelligent}). These works aim at improving the well-known energy-latency trade-off in MEC. However, all state-of-the-art contributions consider static computation-offloading problems, which do not incorporate any adaptation capability in time-varying scenarios (e.g., UE mobility and blocking events), and do not take into account low-latency and ultra-reliability constraints that are typical of many envisaged 5G and 6G applications. Furthermore, to the best of our knowledge, none of the available works has exploited RISs for edge machine learning, with the aim of delivering pervasive and reliable AI to mobile UEs. The WEaaS concept offered by the proposed RISE network paradigm targets providing fiber-like connectivity and reliability, even under intermittent blockages, by the proactive provisioning of communication resources and back-up link generation via the plurality of  available RISs.

\subsection{Environment Awareness and Distributed Intelligence}
The PBA concept relies on highly localized service provisioning, which in turn requires accurate and timely knowledge of the UE locations and the signal propagation conditions, in order to properly control the RISs. It is the vision of the RISE network paradigm that RISs can serve both as users of location and sensing information, as well as a technical enabler~\cite{wymeersch2019radio}. 
The use of RISs for localization has received very limited treatment in the literature so far. In the field of passive detection and mapping, it was shown that an RIS can be used to localize non-cooperative objects at very precise scales in reverberating media~\cite{del2018precise}, and proof-of-concept demonstrations were also performed using WiFi chipsets in indoor environments. 
In~\cite{WD20}, a localization-oriented control was proposed that considers down-selection, activation, and control of phases of the most informative RISs configured as passive reflectors, while limiting the risk of self interference. 

There is a clear need for localization-oriented architectures and control methods, which could operate jointly at the RISs and system levels, while taking maximum benefits from highly directional RIS operations, overall channel reconfigurability, and advanced processing tools (e.g., decentralized/federated machine learning,  multi-objective optimization). Similar to the \textit{communication-boosted areas}, \textit{localization-boosted areas} will guarantee high performance for localization and other sensing KPIs. 
To achieve this, suitable algorithms for localization (location and optionally, attitude estimation) as well as active and passive sensing should also be developed, which relax synchronization and overhead requirements inherent to massive-antenna-based active localization, including most advanced cooperative approaches. These lightweight algorithms will benefit from controlled RIS-enabled multipath profiles in both near- and far-field regimes, to enable high and context-adjustable accuracy and service continuity, while 
maintaining acceptable computational complexity (e.g., for data association).

In addition, the vision of the RISE network paradigm is to also provide distributed situational awareness and intelligence. This refers to being aware of where UEs and passive objects and people are, and what is their history and predicted future, by learning mobility habits and even possibly, checking the consistency of their claimed trajectory (e.g., for trust assessment). This intelligence is part of the proposed overall architecture and will be also used to make proactive decisions at the protocol level in support of the highly focalized PBAs.

\section{Research Roadmap and Conclusion}
Reconfigurable, intelligent, and sustainable environments --enabled by the adoption of RISs-- conceive a novel wireless connectivity paradigm for future 6G networks, while disclosing unprecedented scientific and technological challenges. In this article, we have introduced the RISE view that unveils two main key concepts: the \textit{WEaaS} and the spatio-temporally focused \textit{PBAs}. In particular, we have: $i$) identified the need for revisiting, extending, and harmonizing RIS-based and conventional EM wave propagation by devising realistic channel models, $ii$) discussed the fundamental limits of RIS-empowered networks in terms of connectivity, localization, sensing, and sustainability, $iii$) detailed the definition of network mechanisms incorporating multiple in-band and out-band RIS control channels and related RIS-empowered strategies enabling agility with low re-deployment costs, $iv$) discussed innovative channel estimation and sensing schemes and protocols that blend together high-capacity connectivity, energy efficiency, controlled EMF exposure, and tunable localization accuracy, $v$) highlighted energy sustainable RIS hardware and building blocks showing the architectural integration of RISE into the existing O-RAN-compliant framework, and $vi$) provided practical application scenarios and use cases, where RIS-empowered connectivity is particularly relevant and beneficial through adequate KPIs.

\bibliographystyle{IEEEtran}
\bibliography{references}

\begin{IEEEbiographynophoto}{Emilio Calvanese Strinati} is the 6G Future Technologies Director at  at the French Atomic
Energy Commission's Electronics and Information Technologies Laboratory, Minatec Campus, Grenoble, France. 
\end{IEEEbiographynophoto}
\vskip -2\baselineskip plus -1fil

\begin{IEEEbiographynophoto}{George C. Alexandropoulos} is an Assistant Professor 
with the Department of Informatics and Telecommunications, National and Kapodistrian University of Athens, Greece. His research interests span the general areas of algorithmic design and performance analysis for wireless networks with emphasis on multi-antenna transceiver hardware architectures, reconfigurable metasurfaces, and distributed machine learning algorithms. 
\end{IEEEbiographynophoto}
\vskip -2\baselineskip plus -1fil

\begin{IEEEbiographynophoto}{Henk Wymeersch}
is a Professor 
with the Department of Electrical Engineering at Chalmers University of Technology, Sweden. 
\end{IEEEbiographynophoto}
\vskip -2\baselineskip plus -1fil

\begin{IEEEbiographynophoto}{Beno\^{i}t Denis} is a senior researcher with the 
Department of Wireless Technologies 
at CEA-Leti, Grenoble, France.
\end{IEEEbiographynophoto}
\vskip -2\baselineskip plus -1fil

\begin{IEEEbiographynophoto}{Vincenzo Sciancalepore} is a Principal Researcher at NEC Laboratories Europe, Heidelberg, Germany. 
\end{IEEEbiographynophoto}
\vskip -2\baselineskip plus -1fil

\begin{IEEEbiographynophoto}{Raffaele D'Errico} is a senior researcher with the 
Department of Wireless Technologies 
at CEA-Leti, Grenoble, France.
\end{IEEEbiographynophoto}
\vskip -2\baselineskip plus -1fil

\begin{IEEEbiographynophoto}{Antonio Clemente} is a senior researcher with the 
Department of Wireless Technologies 
at CEA-Leti, Grenoble, France.
\end{IEEEbiographynophoto}
\vskip -2\baselineskip plus -1fil

\begin{IEEEbiographynophoto}{Dinh-Thuy Phan-Huy} is a senior researcher with the Technology \& Global Innovation, Orange Labs Networks, Ch\^{a}tillon, France. 
\end{IEEEbiographynophoto}
\vskip -2\baselineskip plus -1fil

\begin{IEEEbiographynophoto}{Elisabeth De Carvalho} is a Professor at Aalborg University, Denmark.
\end{IEEEbiographynophoto}
\vskip -2\baselineskip plus -1fil

\begin{IEEEbiographynophoto}{Petar Popovski} is a Professor at Aalborg University, Denmark. He is a Fellow of IEEE and is currently a Member at Large on the Board of Governors in IEEE Communication Society. 
\end{IEEEbiographynophoto}

\end{document}